\documentclass[useAMS,usenatbib]{mn2e}
\usepackage{epsfig}
\usepackage{color} 
\usepackage{natbib}
\usepackage{deluxetable}
\usepackage{amssymb}
\usepackage{rotating}
\usepackage{graphicx}
\usepackage[colorlinks=true,citecolor=blue]{hyperref}

\title[Intranight optical variability of radio-quiet WLQs]
{Intranight Optical Variability of Radio-Quiet Weak Emission Line Quasars-IV}
\author[Parveen Kumar, Hum Chand and Gopal-Krishna] {{Parveen Kumar$^{1}$\thanks{E-mail: parveen@aries.res.in (PK);
      hum@aries.res.in (HC); gopaltani@gmail.com (GK)},
    Hum Chand$^{1}$, Gopal-Krishna$^{2}$} \\$^{1}$ Aryabhatta Research
  Institute of Observational Sciences (ARIES), Manora Peak, Nainital,
  263002 India,\\ $^{2}$Centre for Excellence in Basic Sciences (CBS), University of Mumbai campus (Kalina), Mumbai 400098, India
}

\begin{document}
\date{Accepted ---. Received ---; in original form ---}

\pagerange{\pageref{firstpage}--\pageref{lastpage}} \pubyear{2012}

\maketitle

\label{firstpage}
\begin{abstract}
We report an extension of our program to search for radio-quiet BL Lac candidates using intra-night 
optical variability (INOV) as a probe. The present INOV observations cover a well-defined 
representative set of 10 
`radio-quiet weak-emission-line quasars' (RQWLQs), selected from a newly published 
sample of 46 such sources, derived from the Sloan Digital Sky Survey (Data release 7). Intra-night 
CCD monitoring of the 10 RQWLQs was carried out in 18 sessions lasting at least 3.5 hours. 
For each session, differential light curves (DLCs) of the target RQWLQ were derived relative to two 
steady comparison stars monitored simultaneously. Combining these new data with those 
already published by us for 15 RQWLQs monitored in 30 sessions, we estimate an INOV duty cycle of 
$\sim 3\%$ for the RQWLQs, which appears inconsistent with BL Lacs. However, the observed INOV events 
(which occurred in just two of the sessions) are strong (with a fractional variability amplitude
$\psi >$ 10\%), hence blazar-like. We briefly 
point out the prospects of an appreciable rise in the estimated INOV duty cycle for RQWLQs with a
relatively modest increase in sensitivity for monitoring these rather faint objects.

\end{abstract}


\section{Introduction} 
Weak-line-quasars (WLQs), a rare subset of the quasar population,
continue to be an enigma, in-spite of the substantial observational
and theoretical effort invested in probing their nature
(e.g.,~\citet[][] {Plotkin2015ApJ...805..123P}: hereinafter
P15;~\citet[][]{Meusinger2014A&A...568A.114M}: hereinafter
MB14). Exceptional weakness, even absence of emission lines,
particularly in the rest-frame UV spectrum, is their principal
abnormality vis-a-vis normal quasars (e.g., MB14; P15), as
underscored initially by the discoveries of the WLQs: PG 1407+265 at z
= 0.94 ~\citep{McDowell1995ApJ...450..585M} and SDSS
J153259.96-003944.1 at z = 4.67~\citep{Fan1999ApJ...526L..57F}.  

Since then over a hundred of WLQs have been found, mainly using the SDSS
survey ~\citep{York2000AJ....120.1579Y}. Basically, these findings
have given rise to two possible scenarios: (i) WLQs are (predominantly
beamed) BL Lacs whose radiation is uncharacteristically weak in the
radio band, or (ii) they are (unbeamed) quasars with an exceptionally
weak broad emission-line region. While, some rare representatives of
the first scenario may still be discovered among WLQs, the weight of
evidence has steadily shifted towards the second alternative which
appears to be the norm. This inference for WLQs is based on several
observables, such as radio-loudness, optical polarization and
continuum flux variability, all of which are found to be distinctly
milder than those typical of BL Lacs (e.g.,P15 and references therein;
MB14). Furthermore, the rest-frame optical-UV broad-band spectra of
WLQs are mostly found to be matching those of radio-quiet quasars
(e.g.,~\citet{Lane2011ApJ...743..163L};
~\citet{DiamondStanic2009ApJ...699..782D};
see~\citet{Shemmer2009ApJ...696..580S} for a similar inference based
on the X-ray spectra).  Likewise, recent optical polarimetric surveys
of WLQs
~\citep[e.g.,][]{DiamondStanic2009ApJ...699..782D,Heidt2011A&A...529A.162H}
have failed to reveal any robust example of radio-quiet BL Lac, in
accord with earlier findings
~\citep[e.g.,][]{Stocke1990ApJ...348..141S,Jannuzi1994ApJ...428..130J,Smith2007ApJ...663..118S}.

However, despite these negative indications, the first alternative is
not entirely precluded and the possibility remains that at least a
tiny population of radio-quiet BL Lacs may be lurking among WLQs
\citep[e.g.,][MB14]{Londish2004MNRAS.352..903L,Collinge2005AJ....129.2542C,Shemmer2006ApJ...644...86S,Wu2012ApJ...747...10W}.
Intensive searches aimed at picking any such exotic BL Lacs hold
considerable astrophysical interest, since the discovery of even a
single radio-quiet BL Lac would challenge the standard paradigm which
posits that the jets of blazars (of which BL Lacs are a subset) emit
predominantly synchrotron radiation over the (rest-frame)
radio-to-infrared/optical waveband and their entire radiation appears
predominantly relativistically
beamed~\citep[e.g.,][]{Blandford1978PhyS...17..265B,Urry1995PASP..107..803U,Antonucci2012A&AT...27..557A}. Here
it is interesting to recall that although, compared to BL Lacs, RQWLQs
are found to display much milder optical variability on
month/year-like time scale (see, P15; MB14 and references therein), a
few striking exceptions have been reported where a blazar-like large
optical variability was observed on month/year-like time scale,
betraying the presence of relativistically beamed synchrotron
emission. Examples in these RQWLQs are PG 1407+265
at z = 0.94 ~\citep{Blundell2003ApJ...591L.103B} and J153259.96-003944
at z = 4.67 ~\citep{Stalin2005MNRAS.359.1022S}. In PG 1407+265 there
is indeed evidence that a relativistically beamed nonthermal jet
appears intermittently in the radio/X-ray bands
~\citep{Blundell2003ApJ...591L.103B, Gallo2006MNRAS.365..960G}. It may
be recalled that weak parsec-scale relativistic jets have been
detected, or inferred to exist in many radio-quiet quasars
(RQQs)\footnote{Radio-loudness is usually parametrized by the ratio
  (R) of flux densities at 5 GHz and at 2500\AA~in the rest-frame, and
  R $<$ 10 for radio-quiet quasars~\citep[e.g. see,
  ][]{Kellermann1989AJ.....98.1195K}.}, based on radio imaging and
continuum flux variability on month/year-like time scale
\citep[e.g.,][]{Ulvestad2005ApJ...621..123U,Barvainis2005ApJ...618..108B,Blundell1998MNRAS.299..165B,
  Kellermann1994AJ....108.1163K,Czerny2008MNRAS.386.1557C}.
Therefore, it would not be too surprising if weak relativistic jets
were often present even in the subset of RQQs whose members exhibit
uncharacteristically weak emission lines in the UV/optical (i.e.,
RQWLQs). A small fraction of such relativistic jets, oriented close to
the line of sight, would then appear Doppler boosted, as indeed
inferred, e.g., for the RQWLQ PG 1407+265 (see above).

As already discussed widely in the literature, rapid optical
variability on hour-like time scale, termed Intra-Night Optical
Variability (INOV), can be a fairly reliable discriminator between the
AGN whose optical jets are relativistically beamed towards us (i.e.,
blazar-like), and their misaligned (hence unbeamed)
counterparts~\citep[e.g.,][and references
  therein]{Goyal2013MNRAS.435.1300G}. Specifically, the AGNs showing
strong INOV($\psi > 3$\%) are nearly always blazars and the duty cycle
of such strong INOV is around 50\% for a monitoring duration of around
4$-$6 hours
~\citep[e.g.,][]{Goyal2013MNRAS.435.1300G,Carini2007AJ....133..303C,Stalin2004MNRAS.350..175S,
  GopalKrishna2003ApJ...586L..25G}.  Since the INOV data on RQWLQs did
not exist, we have attempted to bridge this gap by initiating a
program of intra-night monitoring of RQWLQs. The results obtained so
far under this program are reported in 3 papers
(~\citet{Gopal2013MNRAS.430.1302G}: Paper
I,~\citet{Chand2014MNRAS.441..726C}: Paper
II,~\citet{Kumar2015MNRAS.448.1463K}: Paper III).  Recently, a similar
program has also been undertaken by~\citet{Liu2015A&A...576A...3L}.
Together, these two INOV programs encompass 18 RQWLQs, of which 15
RQWLQs are covered in our program. This admittedly rather limited
dataset has shown that INOV is a rare occurrence among RQWLQs (duty
cycle $\sim$ 5\%), as also found for radio-quiet quasars and
radio lobe-dominated quasars
~\citep[e.g.,][]{Goyal2013MNRAS.435.1300G,Carini2007AJ....133..303C}. However,
the estimate of INOV duty cycle for RQWLQs may likely be revised
upwards once a matching sensitivity is achieved for monitoring these
relatively faint objects (see Paper III). Thus, the main goal of our
ongoing INOV program is two-fold: (i) to characterise the INOV
behaviour of RQWLQs, and (ii) to make a systematic search for any
blazar-like INOV events among RQWLQs, granting that such events might
be quite rare.
 
\section{THE SAMPLE OF RADIO-QUIET WLQs} 
Since a major goal of our program is to characterise the INOV
properties of RQWLQs, it is desirable to monitor RQWLQ samples
selected from different catalogs, given that individual catalogs are
expected to suffer from different sets of systematic and hence rare
objects, such as WLQs, picked up in them may not represent identical
populations.  Spurred by the initial discoveries of a few individual
cases of WLQs, the huge and rapidly growing SDSS database began to be
deployed to make systematic searches for
WLQs~\citep[e.g.,][]{Collinge2005AJ....129.2542C,DiamondStanic2009ApJ...699..782D}.
In Papers I, II, III we reported INOV observations of 15 bona-fide
RQWLQs monitored in 30 sessions, each lasting for minimum 3 hrs
(median duration 4.2 hr). That well-defined set of RQWLQs was drawn from the list
of 86 RQWLQs published by~\citet{Plotkin2010ApJ...721..562P} who had
classified them as ``high-confidence BL Lac candidate", primarily
because the emission line equivalent-widths are small (Wr $<$ 5\AA)
and the 4000\AA ~break, if present, is less than 40\%. Additional
selection criteria imposed by us were: (i) the object should be
brighter than R {$\sim$} 18.5 mag and (ii) its image should not appear
confused/distorted due to a neighboring object. This is specially
relevant for our type of observations which involve taking a sequence
of CCD exposures and then doing aperture photometry to derive the
light-curve of the monitored target, relative to at least two steady
stars seen in the target's vicinity on each CCD frame. Such
``differential light curves" (DLCs) have become the preferred mode
adopted in INOV studies almost universally
~\citep[e.g.,][]{Miller1989Natur.337..627M}. Lastly, we note that each
of the 15 RQWLQs is consistent with zero proper motion, confirming
their extragalactic nature (Paper III).\par

The present set of 10 RQWLQs, for which INOV results are reported
here, was drawn by us from the list of 46 WLQs published recently in
MB14.  In order to select WLQs they employed machine learning data
mining techniques to the huge database of quasars in the SDSS/DR7
pipeline~\citep[DR-7,][]{Abazajian2009ApJS..182..543A}, followed by
manual inspection of the spectra of individual sources. This led them
to a final sample of 365 quasars with weakly detected emission lines
(as a consistency check, MB14 found all these WLQs to have their
counterparts in the SDSS/DR7 quasar catalog of
\citet{Shen2011ApJS..194...45S}).  From this sample, MB14 extracted a
well-defined sub-sample of 46 WLQs, termed as `WLQ-EWS' with the mean
redshift of 1.48, by imposing rest-frame equivalent-width thresholds:
EW(Mg~{\sc ii})$< 11$\AA~and EW(C~{\sc iv})$< 4.8$\AA, which
represent 3$\sigma$ deviations below the mean of the (log-normal) EW
distribution of the corresponding emission line, for their sample of
365 WLQs. Additional selection criteria imposed by us are (i) the
radio-loudness parameter R $<$ 10, or equivalently, a non-detection in
the FIRST survey, which amounts to a somewhat conservative upper limit
of {$\sim$} 1 mJy at 1.4 GHz for point-like
sources~\citep{Becker1995ApJ...450..559B}; (ii) R-magnitude $<$ 18.5
and (iii) a proper motion consistent with zero
~\citep{Monet2003AJ....125..984M}, so that any Galactic objects are
excluded.  Proper motion is found to be zero for each source,
excepting J134052.43$+$074008.1 which has a proper motion (PM) of
$5.66\pm2.24$ milli-arcsec/yr. We consider this to be consistent with
zero proper motion, and treat this source as extragalactic,
particularly in view of the fact that its SDSS spectrum clearly
exhibits the Mg~{\sc ii} emission line (with a rest-frame equivalent
width of $\sim$ 9.7\AA ~and $z \sim 1.077$, see the catalog of
\citet{Shen2011ApJS..194...45S}; also~\citet{Londish2004MNRAS.352..903L}).  Application of these selection
criteria led us to a well-defined set of 12 RQWLQs, after
rejecting J151554.81$+$251334 on account of its location in a crowded
optical field (see above).  Here we report INOV observations of 10 out
of these 12 RQWLQs, the remaining two sources, J001444.02$-$000018.5 and
J232214.72$-$103725.1 fall outside the 8-17 hr right ascension range
covered in the present observations. Particulars of the observed 10
RQWLQs are given in Table \ref{tab:source_info}.  We note that 4 of
them have also been covered earlier in our INOV program (Paper I) and
those 4 sources are marked with an asterisk in Table
\ref{tab:source_info}.
\begin{table*}
\centering
\begin{minipage}{1000mm}
{\scriptsize
\caption{The set of 10 RQWLQs observed in the present study. \label{tab:source_info}}

\begin{tabular}{lccccccc ccc}
\hline
\multicolumn{1}{l}{IAU Name{\footnote{The sources marked by $^{*}$ have also been covered in 
our earlier publication (Paper I).}}} &  R.A.(J2000) & Dec(J2000) &{\it R-mag} &   
$z${\footnote{All the redshifts are from \citet{Hewett2010MNRAS.405.2302H}, except 
for J134052.43$+$074008.1,  \\whose redshift is taken from \citet{Shen2011ApJS..194...45S}.}} & 
R{\footnote{R is ratio of flux densities at 5 GHz and at 2500\AA~in the rest-frame, 
\citep[e.g.,][]{Kellermann1989AJ.....98.1195K};\\ 
ND=non-detection in the FIRST survey (see text).}}
& PM & Telescope{\footnote{DFOT=Devasthal Fast Optical Telescope; ST=Sampurnanand Telescope.}}\\
         & (h m s)      &($ ^{\circ}$ $ ^{\prime}$ $ ^{\prime\prime }$) &  &     & & (msec/yr)& used & \\
 (1)     &(2)             &(3)                             &(4)     &(5) &(6) &(7) & (8)\\
\hline
\multicolumn{5}{l}{}\\
J081250.79$+$522531.0$^{*}$&  08 12 50.80&$+$ 52 25 31.0 & 18.30 &1.1532$\pm$ 0.0011 &2.75&0&ST \\
J083232.37$+$430306.1      &  08 32 32.37&$+$ 43 03 06.1 & 17.95 &1.3136$\pm$ 0.0007&ND&0&DFOT \\
J094726.72$+$443526.5      &  09 47 26.72&$+$ 44 35 26.5 & 18.18 &1.2887$\pm$ 0.0007&ND&0 &DFOT \\
J110539.59$+$315955.6      &  11 05 39.59&$+$ 31 59 55.6 & 18.46 &1.7824$\pm$ 0.0012&ND&0&DFOT \\
J113413.48$+$001042.0      &  11 34 13.48&$+$ 00 10 42.0 & 18.46 &1.4857$\pm$ 0.0007&ND&0&DFOT \\
J124514.04$+$563916.1      &  12 45 14.04&$+$ 56 39 16.1 & 18.47 &0.6139$\pm$ 0.0004&ND&0&DFOT \\
J125219.47$+$264053.9$^{*}$&  12 52 19.47&$+$ 26 40 53.9 & 17.72 &1.2883$\pm$ 0.0007&4.51&0&DFOT \\
J134052.43$+$074008.1      &  13 40 52.43&$+$ 07 40 08.1 & 17.95 &1.0773$\pm$ 0.0004&ND&5.66$\pm$2.24&DFOT \\
J142943.60$+$385932.0$^{*}$&  14 29 43.60&$+$ 38 59 32.0 & 17.56 &0.9279$\pm$ 0.0005&ND&0&DFOT \\
J161245.68$+$511817.3$^{*}$&  16 12 45.68&$+$ 51 18 17.3 & 17.73 &1.5942$\pm$ 0.0010&ND&0&DFOT \\

\hline
\end{tabular}
}
 \end{minipage}
\end{table*} 

\subsection{The Photometric Monitoring}
The monitoring was done in the SDSS r-band using the 1.3-m optical telescope (DFOT 
\footnote {Devasthal Fast Optical Telescope, located near Nainital
  (India)}) ~\citep{Sagar2011Csi...101...8.25}, except for one source,
J081250.79$+$522531.0 which was monitored using the 1.04-m
Sampurnanand Telescope (ST) located at Nainital, India.  Each time, a
given source was monitored for a minimum duration of 3 hours.  Table
\ref{wl:tab_res} provides the log of the monitoring sessions.

The 1.3-m DFOT is a fast beam (f/4) optical telescope with a pointing
accuracy better than $10$ arcsec (rms). It is equipped with a 512k
$\times$ 512k Andor CCD camera having a pixel size of 16 micron and a
plate scale of 0.63 arcsec per pixel.  The CCD covers a field of view
of $\sim$ 5 arcmin on the sky. It is cooled thermo-electrically to
-$90$ degC and is read out at 1 MHz speed. The corresponding system
noise is $6.1$ e- (rms) and the gain is $1.4$ e-/Analog to Digital
Unit (ADU).

 The 1.04-m ST is equipped with a 2k$\times$2k liquid-nitrogen cooled
 CCD camera having square pixels of 24 micron and a plate scale of
 0.37 per pixel. The CCD covers a square field-of-view of about 13
 arcmin on a side.  Operating at 27 kHz, the gain and readout noise of
 the CCD are 10e$^{-}$ per Analog-to-Digital Unit (ADU) and
 5.3e$^{-}$, respectively.

The exposure time for each science frame was set to about $5-7$
minute, yielding a typical SNR above $25-30$. The typical seeing
(FWHM) during our observations is close to 2 arcsec.  Since in the
sample selection, care was taken to ensure the availability of at
least two, but usually more, comparison stars on each CCD frame,
within about 1 mag of the target RQWLQ, it became possible to identify
and discount any comparison star(s) which showed a hint of variability
during the monitoring session.
 
\subsection{The Data Reduction}
\label{wl:sec_data}
The pre-processing of the raw images (bias subtraction, flat-fielding,
cosmic-ray removal and trimming) was done using the standard tasks
available in the Image Reduction and Analysis Facility {\textsc
  IRAF} \footnote{\textsc {Image Reduction and Analysis Facility
    (http://iraf.noao.edu/). }}.  The instrumental magnitudes of the
observed RQWLQs and their chosen comparison stars in the CCD frames
were determined by aperture photometry~\citep{1992ASPC...25..297S,
  1987PASP...99..191S}, using the Dominion Astronomical Observatory
Photometry \textrm{II} (DAOPHOT II algorithm)\footnote{\textsc
  {Dominion Astrophysical Observatory Photometry.}}.  To select the
aperture size (FWHM) for photometry, we first determined the ``seeing"
for each frame by averaging the observed FWHMs of $5$ moderately bright stars
in the frame. We then took the median of these averaged values over
all the frames recorded in the session. The aperture diameter was set
equal to 2 times the median FWHM. \par
 
To derive the Differential Light Curves (DLCs) of the target RQWLQ
monitored in a given session, we selected two steady comparison star
present within each CCD frame, on the basis of their proximity to the
target, both in apparent location and brightness. Particulars of the comparison
stars used for the various sessions are given in Table
~\ref{tab_cdq_comp}.  Note that the $g-r$ color difference for our
`quasar-star' and `star-star' pairs is always $< 1.5$, with a median
value of $0.5$ (Table ~\ref{tab_cdq_comp}, column 7). Analyses by
~\citet{Carini1992AJ....104...15C} and
~\citet{Stalin2004MNRAS.350..175S,Stalin2004JApA...25....1S}, show that for color difference of
this order, the changing atmospheric attenuation during a session
produces a negligible effect on the DLCs.\par
\begin{table*}
\centering
\caption{Basic parameters and observing dates (18 sessions) for the $10$ RQWLQs and their comparison stars.
\label{tab_cdq_comp}}
\scriptsize
\begin{tabular}{ccc ccc c}\\
\hline

{IAU Name} &   Date       &   {R.A.(J2000)} & {Dec.(J2000)}                      & {\it g} & {\it r} & {\it g-r} \\
           &  dd.mm.yyyy    &   (h m s)       &($^\circ$ $^\prime$ $^{\prime\prime}$)   & (mag)   & (mag)   & (mag)     \\
{(1)}      & {(2)}        & {(3)}           & {(4)}                              & {(5)}   & {(6)}   & {(7)}     \\
\hline
\multicolumn{7}{l}{}\\

J081250.79$+$522531.0 &  24.12.2014      &08 12 50.79 &$+$52 25 31.0  &   18.30 &       18.05 &        0.25\\   
S1                    &                  &08 13 29.57 &$+$52 27 56.3  &   20.14 &       18.60 &        1.54\\
S2                    &                  &08 13 20.70 &$+$52 23 27.8  &   18.36 &       17.80 &        0.56\\
J081250.79$+$522531.0 &  25.12.2014      &08 12 50.79 &$+$52 25 31.0  &   18.30 &       18.05 &        0.25\\   
S1                    &                  &08 13 20.70 &$+$52 23 27.8  &   18.36 &       17.80 &        0.56\\
S2                    &                  &08 13 21.90 &$+$52 24 58.8  &   19.23 &       17.81 &        1.42\\
J083232.37$+$430306.1 &  08.11.2015      &08 32 32.37 &$+$43 03 06.1  &   18.12 &       17.95 &        0.17\\
S1                    &                  &08 32 34.56 &$+$43 01 34.9  &   18.57 &       17.08 &        1.49\\
S2                    &                  &08 32 24.30 &$+$43 03 10.4  &   17.54 &       16.94 &        0.60\\
J083232.37$+$430306.1 &  09.11.2015      &08 32 32.37 &$+$43 03 06.1  &   18.12 &       17.95 &        0.17\\
S1                    &                  &08 32 40.67 &$+$43 04 39.3  &   17.96 &       17.54 &        0.42\\
S2                    &                  &08 32 24.30 &$+$43 03 10.4  &   17.54 &       16.94 &        0.60\\
J083232.37$+$430306.1 &  10.11.2015      &08 32 32.37 &$+$43 03 06.1  &   18.12 &       17.95 &        0.17\\
S1                    &                  &08 32 40.67 &$+$43 04 39.3  &   17.96 &       17.54 &        0.42\\
S2                    &                  &08 32 34.56 &$+$43 01 34.9  &   18.57 &       17.08 &        1.49\\
J083232.37$+$430306.1 &  01.02.2016      &08 32 32.37 &$+$43 03 06.1  &   18.12 &       17.95 &        0.17\\
S1                    &                  &08 32 40.67 &$+$43 04 39.3  &   17.96 &       17.54 &        0.42\\
S2                    &                  &08 32 34.56 &$+$43 01 34.9  &   18.57 &       17.08 &        1.49\\
J083232.37$+$430306.1 &  02.02.2016      &08 32 32.37 &$+$43 03 06.1  &   18.12 &       17.95 &        0.17\\
S1                    &                  &08 32 46.22 &$+$43 02 41.7  &   18.93 &       17.48 &        1.45\\
S2                    &                  &08 32 34.56 &$+$43 01 34.9  &   18.57 &       17.08 &        1.49\\
J083232.37$+$430306.1 &  03.02.2016      &08 32 32.37 &$+$43 03 06.1  &   18.12 &       17.95 &        0.17\\
S1                    &                  &08 32 40.67 &$+$43 04 39.3  &   17.96 &       17.54 &        0.42\\
S2                    &                  &08 32 34.56 &$+$43 01 34.9  &   18.57 &       17.08 &        1.49\\
J094726.72$+$443526.5 &  18.12.2015      &09 47 26.72 &$+$44 35 26.5  &   18.23 &       18.17 &        0.06\\
S1                    &                  &09 47 17.61 &$+$44 35 05.6  &   19.35 &       18.14 &        1.21\\
S2                    &                  &09 47 38.95 &$+$44 34 29.1  &   19.06 &       17.76 &        1.30\\
J110539.59$+$315955.6 &  02.02.2016      &11 05 39.59 &$+$31 59 55.6  &   18.64 &       18.48 &        0.16\\
S1                    &                  &11 05 42.26 &$+$32 02 21.8  &   17.94 &       17.40 &        0.54\\
S2                    &                  &11 05 50.05 &$+$32 00 56.9  &   18.61 &       17.23 &        1.38\\
J113413.48$+$001042.0 &  03.04.2016      &11 34 13.48 &$+$00 10 42.0  &   18.72 &       18.44 &        0.28\\
S1                    &                  &11 34 22.50 &$+$00 10 34.5  &   19.16 &       17.75 &        1.41\\
S2                    &                  &11 34 09.65 &$+$00 11 12.9  &   18.04 &       17.64 &        0.40\\
J124514.04$+$563916.1 &  04.02.2016      &12 45 14.04 &$+$56 39 16.1  &   18.57 &       18.44 &        0.13\\
S1                    &                  &12 44 54.50 &$+$56 36 45.3  &   18.70 &       18.19 &        0.51\\
S2                    &                  &12 44 51.84 &$+$56 39 38.4  &   17.65 &       17.31 &        0.34\\
J125219.47$+$264053.9 &  05.04.2016      &12 52 19.47 &$+$26 40 53.9  &   17.94 &       17.70 &        0.24\\
S1                    &                  &12 52 14.26 &$+$26 39 11.5  &   18.43 &       17.15 &        1.28\\
S2                    &                  &12 52 23.82 &$+$26 41 42.6  &   16.71 &       16.43 &        0.28\\
J134052.43$+$074008.1 &  02.04.2016      &13 40 52.43 &$+$07 40 08.1  &   18.08 &       17.95 &        0.13\\
S1                    &                  &13 41 02.94 &$+$07 38 32.9  &   18.20 &       16.71 &        1.49\\
S2                    &                  &13 40 51.00 &$+$07 40 02.6  &   19.07 &       17.72 &        1.35\\
J142943.64$+$385932.2 &  10.05.2016      &14 29 43.64 &$+$38 59 32.2  &   17.56 &       17.55 &        0.01\\
S1                    &                  &14 29 49.95 &$+$39 00 15.6  &   17.30 &       16.54 &        0.76\\
S2                    &                  &14 29 59.94 &$+$39 00 49.8  &   17.50 &       16.13 &        1.37\\
J161245.67$+$511816.9 &  03.04.2016      &16 12 45.67 &$+$51 18 16.9  &   17.93 &       17.76 &        0.17\\
S1                    &                  &16 12 38.59 &$+$51 19 48.4  &   16.50 &       16.14 &        0.36\\
S2                    &                  &16 12 30.03 &$+$51 17 10.4  &   18.85 &       17.39 &        1.46\\
J161245.67$+$511816.9 &  13.04.2016      &16 12 45.67 &$+$51 18 16.9  &   17.93 &       17.76 &        0.17\\
S1                    &                  &16 12 50.46 &$+$51 19 41.1  &   18.36 &       17.71 &        0.65\\
S2                    &                  &16 12 39.38 &$+$51 19 23.5  &   18.76 &       17.68 &        1.08\\
J161245.67$+$511816.9 &  11.05.2016      &16 12 45.67 &$+$51 18 16.9  &   17.93 &       17.76 &        0.17\\
S1                    &                  &16 12 30.03 &$+$51 17 10.4  &   18.85 &       17.39 &        1.46\\
S2                    &                  &16 12 30.58 &$+$51 20 07.3  &   17.97 &       17.37 &        0.60\\
\hline
\end{tabular}
\end{table*}

\section{STATISTICAL ANALYSIS}  
C-statistic~\citep[e.g.,][]{1997AJ....114..565J} is the most commonly used and the one-way analysis 
of variance (ANOVA)~\citep{Diego2010AJ....139.1269D} the most powerful test for verifying 
the presence of variability in a DLC. However, we did not employ either of these tests 
because,~\citet{Diego2010AJ....139.1269D} has questioned the validity of the C-test by 
arguing that the C-statistics does not have a Gaussian distribution and the commonly used
critical value of 2.567 is too conservative. 
On the other hand, the ANOVA test requires a rather large number of data 
points in the DLC, so as to have several points within each sub-group used for the analysis. 
This is not feasible for our DLCs which typically have no more than about 30$-$45 
data points. So, we have instead used the \emph{F$-$test} which is based on the ratio of 
variances, {\it F }$= variance(observed)/variance(expected)$~\citep{Diego2010AJ....139.1269D,Villforth2010ApJ...723..737V},
with its two versions : (i) the standard {\it F$-$test} 
(hereafter $F^{\eta}-$test,~\citet{Goyal2012A&A...544A..37G}) and 
(ii) scaled $F-$test (hereafter $F^{\kappa}-$test,~\citet{Joshi2011MNRAS.412.2717J}).  
The $F^{\kappa}-$test is preferred when the magnitude difference between the object and 
comparison stars is large ~\citep{Joshi2011MNRAS.412.2717J}. Onward Paper II, we have only 
been using the $F^{\eta}-$test because our objects are generally quite comparable in 
brightness to their available comparison stars. An additional gain from the use of the 
$F^{\eta}-$test is that we can directly compare our INOV results with those deduced for 
other major AGN classes ~\citep{Goyal2013MNRAS.435.1300G}. An important point to keep in 
mind while applying the statistical tests is that the photometric errors on individual 
data points in a given DLC, as returned by the algorithms in the IRAF and DAOPHOT softwares 
are normally underestimated by the 
factor $\eta$ which ranges between $1.3$ and $1.75$, as estimated in independent 
studies~\citep[e.g.,][]
{1995MNRAS.274..701G, 1999MNRAS.309..803G, Sagar2004MNRAS.348..176S, 
  Stalin2004JApA...25....1S, Bachev2005MNRAS.358..774B}. Recently,
using a large sample,~\citet{Goyal2013MNRAS.435.1300G} estimated the best-fit value of $\eta$ to be
$1.5$, which is adopted here. Thus, the $F^{\eta}-$ statistics can be expressed as:  
\begin{equation} 
 \label{eq.fetest}
F_{1}^{\eta} = \frac{\sigma^{2}_{(q-s1)}}
{ \eta^2 \langle \sigma_{q-s1}^2 \rangle}, \nonumber  \\
\hspace{0.2cm} F_{2}^{\eta} = \frac{\sigma^{2}_{(q-s2)}}
{ \eta^2 \langle \sigma_{q-s2}^2 \rangle},\nonumber  \\
\hspace{0.2cm} F_{s1-s2}^{\eta} = \frac{\sigma^{2}_{(s1-s2)}}
{ \eta^2 \langle \sigma_{s1-s2}^2 \rangle}
\end{equation}
where $\sigma^{2}_{(q-s1)}$, $\sigma^{2}_{(q-s2)}$ and
$\sigma^{2}_{(s1-s2)}$ are the variances of the `quasar-star1',
`quasar-star2' and `star1-star2' DLCs and $\langle \sigma_{q-s1}^2
\rangle=\sum_\mathbf{i=0}^{N}\sigma^2_{i,err}(q-s1)/N$, $\langle \sigma_{q-s2}^2 \rangle$ and
$\langle \sigma_{s1-s2}^2 \rangle$ are the mean square (formal) rms
errors of the individual data points in the `quasar-star1', `quasar-star2'
and `star1-star2' DLCs, respectively. $\eta$ is the scaling factor (=$1.5$) as mentioned above.

The $F^{\eta}$-test is applied to a given DLC (say, q-s) by calculating the $F$ value using Eq.~\ref{eq.fetest}, 
and then comparing it with the critical value, $F^{(\alpha)}_{\nu_{qs}}$, 
where $\alpha$ is the significance level set for the test, and 
$\nu_{qs}$ is the degree of freedom (N $-$ 1) of the `quasar-star' DLC. 
The two values we have set for the significance level are $\alpha=$ 0.01 and 0.05, which 
correspond to confidence levels of greater than 99 and 95 per cent, respectively. If the computed $F$ 
value exceeds the corresponding critical value $F_{c}$, the null hypothesis (i.e., no variability) 
is discarded to the respective level of confidence. Thus, a RQWLQ is marked as \emph{variable} 
(`V') for a given session, if the computed $F$-values for both its DLCs are $\ge F_{c}(0.99)$, 
which corresponds to a confidence level $\ge 99$ per cent, and is termed \emph{non- variable} 
(`NV') if either of the two DLCs is found to have an $F$-value $\le F_{c}(0.95)$. The remaining 
cases are classified as \emph{probably variable} (`PV').

The inferred INOV status of the DLCs of each RQWLQ, relative to the two comparison stars, 
are presented in Table~\ref{wl:tab_res} for each monitoring session. 
In the first 4 columns, we list the name of the RQWLQ, 
the date and duration (T) of its monitoring and the number of data points (N) (which is
the same for both DLCs of the RQWLQ). 
The next two columns list the computed $F$-values for the two
DLCs and their INOV status, based on the $F^{\eta}-$test.
Column $7$ gives our averaged photometric error $\sigma_{i,err}(q-s)$ of the data points in 
the two `quasar$-$star' DLCs.
Typically, it lies between 0.02 and 0.06 mag (without the $\eta$ scaling mentioned above).

\section{DISCUSSION AND CONCLUSIONS}
\label{wl:sec_dis}
This paper extends our program aimed at the first systematic
characterisation of the INOV properties of ``radio-quiet weak-line
quasars'' (RQWLQs). This program began about four years ago and
differential light-curves (DLCs) of 15 bona-fide RQWLQs, monitored in
30 sessions of $\ge$ 3 hours each, have been reported in 3 papers, as
summarized in Papers III. To derive the differential light curves, we
basically followed the analysis procedure and statistical test, very
similar to those adopted in~\citet{Goyal2013MNRAS.435.1300G} for
determining the INOV properties of AGN of different classes. Based on
the 30 monitoring sessions, of which two showed INOV (for the RQWLQs
J090843.25$+$285229.8 and J140710.26$+$241853.6), an INOV duty cycle
of $\sim$ 5\% was estimated for RQWLQs (Paper III). The procedure
followed for this has been outlined in Paper I~\citep[see,
  also,][]{Romero1999A&AS..135..477R}. If we now add to these 30 DLCs
the 18 DLCs presented here (none of which shows a significant INOV),
the estimate of INOV duty cycle drops to $\sim 3.2\%$ which is
comparable to the INOV duty cycles found for radio-quiet quasars and
radio lobe-dominated quasars, also by applying the
$F^{\eta}-$test~\citep{Goyal2013MNRAS.435.1300G}.
However, we note that the estimated INOV duty cycle for RQWLQs is probably an
underestimate since the present set of 10 RQWLQs has been selected on the criterion
of weakly detected emission lines, instead of a featureless (i.e., BL Lac type)
optical/UV spectrum (Sect. 2; MB14).
 
Another pertinent point is that the presence of weak emission lines in
the SDSS spectra of all the 10 RQWLQs monitored here (Sect. 2) does
not preclude a BL Lac nature for at least some of them. As pointed out
by ~\citet{Collinge2005AJ....129.2542C}, the faint emission lines
could even be contributed by H~{\sc ii} regions in the host galaxy.
Secondly, BL Lac signatures can be transitory; even the prototypical
BL Lac itself has shown emission lines in its optical spectrum, at a
level above the standard threshold defined for BL Lacs
\citep{Vermeulen1996AJ....111.1013V}. The non-detection of optical
linear polarization above $2-3$\% level is another well known argument
used to discount a blazar interpretation of RQWLQs ~\citep[e.g.,]
[Section
  1]{Stocke1990ApJ...348..141S,Jannuzi1994ApJ...428..130J,Smith2007ApJ...663..118S}.
However, again, this by itself does not exclude the presence of a tiny
subset of BL Lacs among RQWLQs, given that a large polarization
variability is a hallmark of BL Lacs
~\citep{Angel1980ARA&A..18..321A}. One may recall some independent
studies
~\citep[e.g.,][]{Fugmann1988A&A...205...86F,Jannuzi1989BAAS...21..717J}
demonstrating that there is a good chance ($\sim$ 40\%) that at a
given epoch, even a bona-fide blazar may not show a high optical
polarization (i.e., above 3$-$4\% level). All these considerations
provide an impetus to extend searches for the elusive radio-quiet (or
radio-weak) BL Lac objects, since their existence can be a crucial
ingredient to development of a comprehensive theoretical understanding
of the AGN jets. Intra-night monitoring offers a useful practical tool
to address this issue and even a modest enhancement in the sensitivity
is likely to boost the efficacy of this approach substantially.

\begin{table*}
\centering
\begin{minipage}{500mm}
{\small
\caption{Observational details and INOV results for the sample of 10 RQWLQs monitored in 18 sessions.}
\label{wl:tab_res}
\begin{tabular}{@{}ccc c rrr rrr cc@{}}
\hline \multicolumn{1}{c}{RQWLQ} &{Date} &{T} &{N} 
&\multicolumn{1}{c}{F-test values} 
&\multicolumn{1}{c}{INOV status{\footnote{V=variable, i.e., confidence level
$\ge 0.99$; PV=probable variable, i.e., $0.95-0.99$ confidence level;
\\NV=non-variable, i.e., confidence level $< 0.95$.
Variability status identifiers based on\\ quasar-star1 and quasar-star2
pairs are separated by a comma.}}}
&{$\sqrt { \langle \sigma^2_{i,err} \rangle}$}\\
& dd.mm.yyyy& 
hr & &{$F_1^{\eta}$},{$F_2^{\eta}$}
&F$_{\eta}$-test &(q-s)\\
(1)&(2) &(3) &(4) &(5)&(6)
&(7)\\ \hline
J081250.79$+$522531.0 &24.11.2014& 5.10& 32& 1.35,1.57 & NV, NV&0.04\\
J081250.79$+$522531.0 &25.11.2014& 5.45& 45& 0.23,0.33 & NV, NV&0.04\\ 
J083232.37$+$430306.1 &08.11.2015& 3.43& 31& 0.13,0.15 & NV, NV&0.04\\
J083232.37$+$430306.1 &09.11.2015& 3.66& 31& 0.35,0.16 & NV, NV&0.03\\
J083232.37$+$430306.1 &10.11.2015& 3.23& 27& 0.27,0.12 & NV, NV&0.03\\
J083232.37$+$430306.1 &01.02.2016& 4.26& 33& 0.24,0.17 & NV, NV&0.04\\
J083232.37$+$430306.1 &02.02.2016& 4.30& 33& 0.36,0.39 & NV, NV&0.05\\
J083232.37$+$430306.1 &03.02.2016& 4.15& 26& 0.39,0.29 & NV, NV&0.04\\ 
J094726.72$+$443526.5 &18.12.2015& 4.56& 39& 0.18,0.28 & NV, NV&0.04\\ 
J110539.59$+$315955.6 &02.02.2016& 4.65& 37& 0.35,0.38 & NV, NV&0.06\\
J113413.48$+$001042.0 &03.04.2016& 3.34& 26& 0.25,0.24 & NV, NV&0.06\\
J124514.04$+$563916.1 &04.02.2016& 3.68& 28& 0.16,0.20 & NV, NV&0.05\\ 
J125219.47$+$264053.9 &05.04.2016& 4.52& 31& 0.16,0.19 & NV, NV&0.06\\
J134052.43$+$074008.1 &02.04.2016& 4.69& 34& 0.66,0.54 & NV, NV&0.05\\
J142943.64$+$385932.2 &10.05.2016& 5.68& 45& 0.20,0.24 & NV, NV&0.05\\
J161245.68$+$511817.3 &03.04.2016& 3.73& 25& 0.52,0.33 & NV, NV&0.04\\
J161245.68$+$511817.3 &13.04.2016& 3.85& 32& 0.27,0.21 & NV, NV&0.05\\
J161245.68$+$511817.3 &11.05.2016& 3.36& 29& 0.13,0.22 & NV, NV&0.05\\
\hline
\end{tabular}
}
\end{minipage}
\end{table*}

\begin{figure*} 
\centering
\psfig{figure=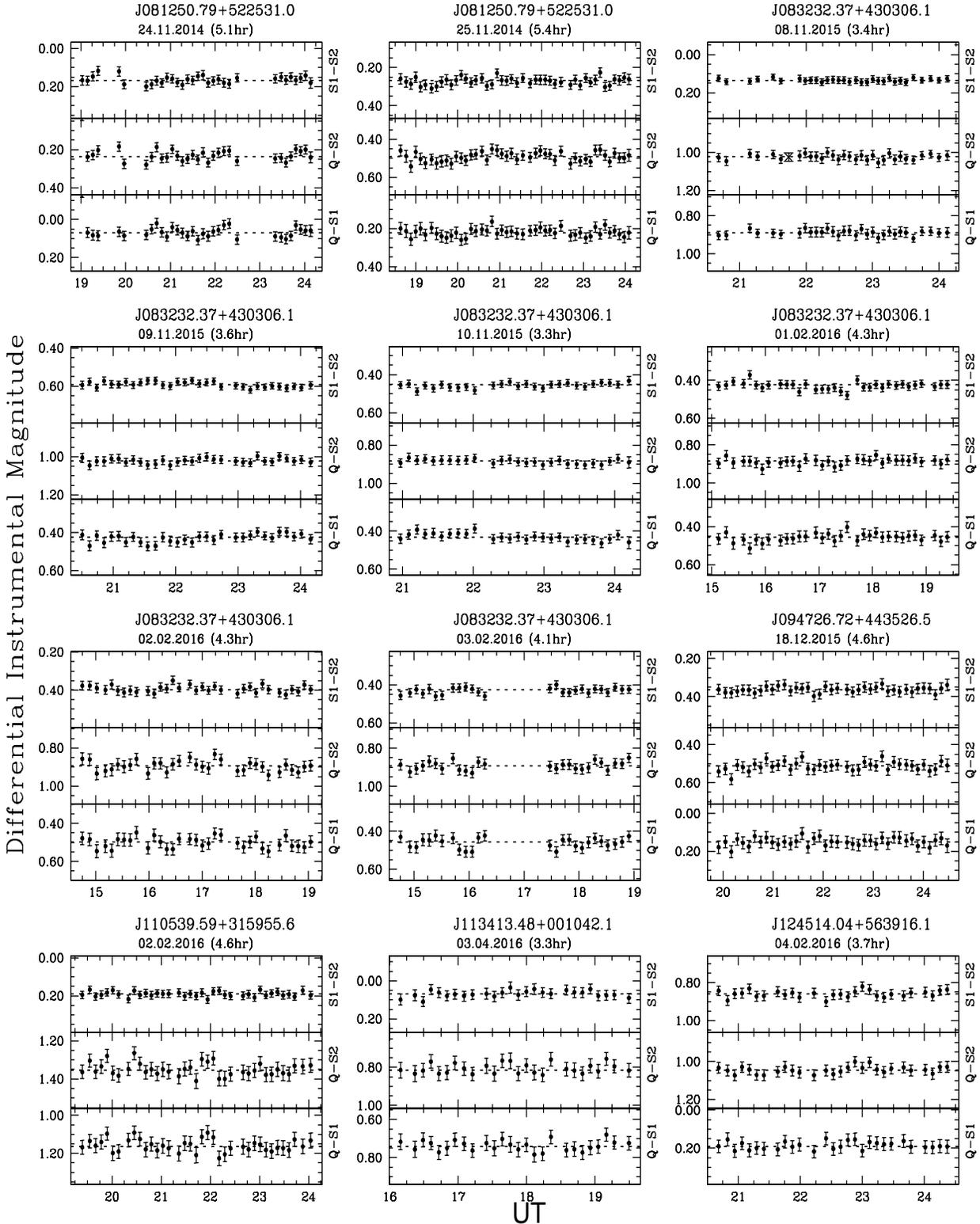,height=20.8cm,width=18.0cm,angle=00,bbllx=20bp,bblly=142bp,bburx=580bp,bbury=711bp,clip=true}
\vspace{-0.5cm}
\caption[]{Differential light curves (DLCs) in R-band, for our sample of $10$ RQWLQs. 
The name of the RQWLQ along with the date and duration of its monitoring session are given 
at the top of each panel. In each panel, the upper DLC is derived using the two `non-varying'
comparison stars, while the lower two DLCs are the `quasar-star' DLCs, as defined in the 
labels on the right side. Any likely outlier points (at $> 3\sigma$) in the DLCs are marked 
with crosses and such points are excluded from the statistical analysis. The number of such points does not
exceed to for any of the DLCs.}
\label{fig:lurve1}
 \end{figure*}

\begin{figure*} 
\centering
\psfig{figure=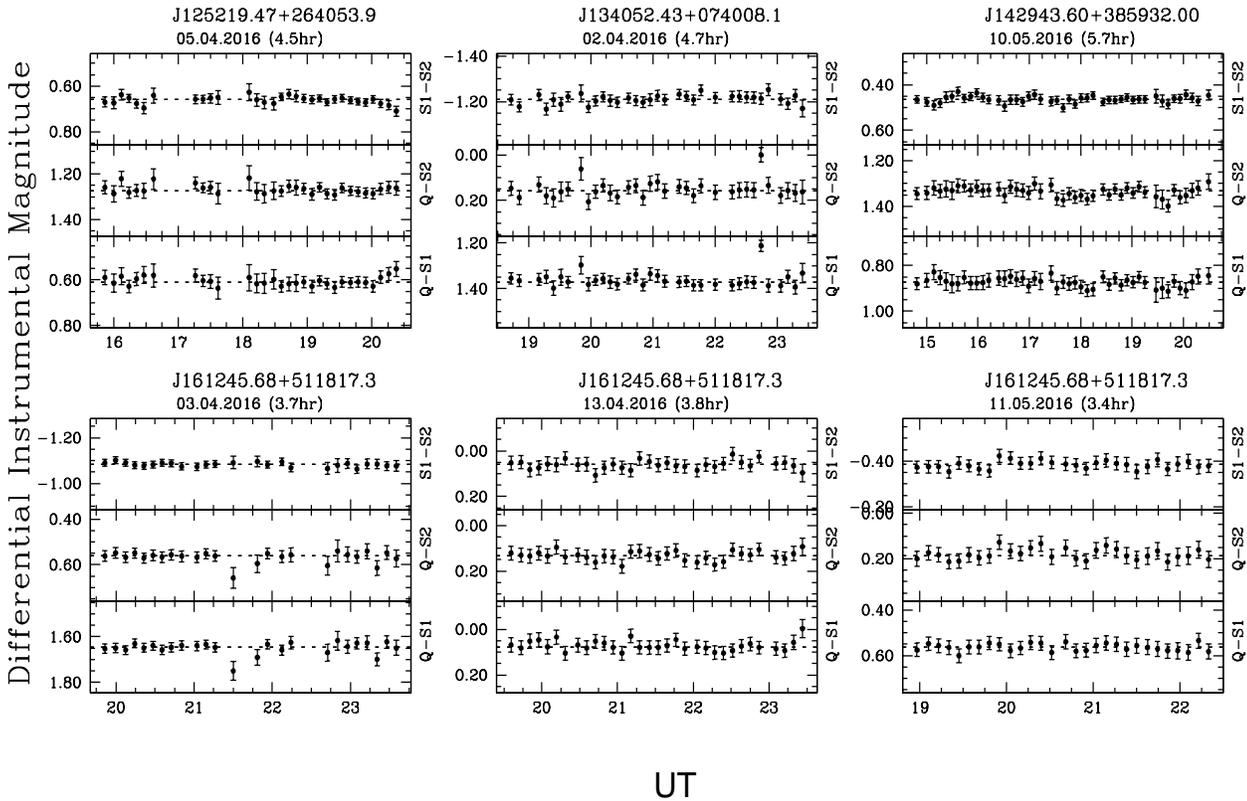,height=11.0cm,width=20.0cm,angle=00,bbllx=20bp,bblly=394bp,bburx=636bp,bbury=711bp,clip=true}
\caption[]{Same as Figure~\ref{fig:lurve1}.}
\label{fig:lurve2}
 \end{figure*}
\section*{Acknowledgments}
G-K thanks the National Academy of Sciences, India for the award of a NASI Senior 
Scientist Platinum Jubilee Fellowship. \par

\bibliography{references}
\label{lastpage}
\end{document}